\documentclass[twocolumn,showpacs,preprintnumbers,amsmath,amssymb]{revtex4}
\usepackage{graphicx}
\usepackage{dcolumn}
\usepackage{bm}

\begin{document}

\title{Unification of the standard and gradient theories of phase transition}

\author{B. I. Lev and A. G. Zagorodny}

\affiliation{Bogolyubov Institute for Theoretical Physics, 
NAS Ukraine, Metrolohichna 14-b, Kyiv 03680, Ukraine}
\date{\today}

\begin{abstract}

We show, that the standard model of phase transition can be unified 
with the gradient model of phase transitions using the description 
in  terms of  the gradient of order parameter. The generalization of 
the gradient theory of phase transitions with regard to the fourth 
power of the order parameter and its gradient is proposed. Such 
generalization makes it possible to described wide class of phase
transitions within a unified approach. In particular it is consistent
with the nonlinear models, that can be used to describe a phase 
transition with the formation of spatially inhomogeneous distribution 
of the order parameter. Typical examples of such structures (with or
without defects) are considered. We  show that formation of spatially
inhomogeneous distributions of the order parameter in the course of a
phase transitions is a characteristic feature of many nonlinear 
models of phase transitions. 
\end{abstract}

\maketitle

Withing the context of the general theory of phase transitions a system 
treated as a continuous medium is assumed to have a ground state which 
can always be described in terms of the order parameter. Such order 
parameter can have various geometrical presentations, for example, 
a scalar field in the case of condensed matter \cite{Lan},a fundamental 
scalar field in the quantum field theory \cite{Lin}, a magnetization 
vector in the theory of magnetism \cite{Lupe}, a second-rank tensor 
in the liquid crystal theory \cite{Gen},etc. 

To introduce the order parameter that determines a stable state
of condensed matter, we have to consider possible deformations of the 
distribution of the field, in particular,  the disordered
configuration of the ground state. 

The phase transition associated with the system with broken continuous 
symmetry can be described in terms of the relevant order parameter. 
In particular, according to the Landau theory the free energy density
can be presented in terms of the order parameter as given by 
\begin{equation}
 f=\left(\mathbf{\nabla}\varphi(\mathbf{r})\right)^{2}+W(\varphi(\mathbf{r}))
\end{equation} 
where $\varphi(\mathbf{r})$ is the order parameter and $W(\varphi(\mathbf{r}))$ 
describes the order parameter dependence of the free energy that is assumed 
to be known.

In the well-known standard model of phase transitions 
\begin{equation}
 W(\varphi(\mathbf{r}))=\frac{1}{2}a\varphi^{2}(\mathbf{r})+\frac{1}{4}b \varphi^{4}(\mathbf{r}).
\end{equation} 
With the dimensionless variable $\varphi^{2}(\mathbf{r})=\frac{b\varphi^{2}(\mathbf{r})}{a}$ 
being introduced the standard dimensionless free energy reduced to the form given by 
\begin{equation}
 f=l^{2}\left(\mathbf{\nabla}\varphi(\mathbf{r})\right)^{2}+
(\left(\varphi(\mathbf{r})\right)^{2}-1)^{2}
\end{equation}
with the potential being written in the standard form and $l^{2}=\frac{2}{a}$
being the characteristic length. Making use of this expression for the 
free energy, we can find the spatial distribution of the order parameter 
and thus, describe the properties of the new states which can be formed 
after the phase transition. 

It should be noted that in the case of system with the gradient of the 
order parameter, it looks reasonable to introduce into the  free energy 
functional the term responsible for possible interaction between the order 
parameter and its gradient. This coupling can regularize possible 
perturbations of the order parameter and thus confine the  spatially 
inhomogeneous state of the system.

On the other hand, singular perturbation models, involving higher order
of the the spatial derivations have provided a new insight on the role of 
additional physical features of the system under consideration, on the 
details of phase transitions, in particular on the way how to describe 
the contribution of the surface energy. In this case restricting the 
accuracy of the free energy functional by the first order derivatives 
of the order parameter leads to the solution with the homogeneous 
distribution of the order parameter only.

The main idea of the present contribution is to generalize the phenomenological 
theory of phase transitions by introducing the second-order spatial derivatives
into the order parameter dependence of the free energy. Such generalization
is reasonable in view both mathematical and physical arguments. From the 
mathematical point of view, our argument is that the order parameter can be
treated as a vector quantity and thus it is possible to rewrite the free energy (1)
in the form 
\begin{equation}
 f=l^{2}\left(\mathbf{\nabla}\mathbf{\varphi}(\mathbf{r})\right)^{2}+(\left(\left|\vec{\varphi}(\mathbf{r})\right|\right)^{2}-1)^{2}
\end{equation}
where the order parameter is a vector function. Such problem arises 
if we assume that the order parameter $\varphi(\mathbf{r}$ can be presented 
is in term of a gradient of some other scalar function $ \mathbf{\nabla}u(\mathbf{r}$. 
This assumption leads to the known presentation of the free energy in the Aviles-Giga 
form \cite{Avi},\cite{Gig},i.e.
\begin{equation} 
f=l^{2}\left(\left|\mathbf{\nabla}\mathbf{\nabla}u(\mathbf{r})\right|\right)^{2}+
(\left(\left|\mathbf{\nabla}u(\mathbf{r}\right|\right)^{2}-1)^{2}
\end{equation}
As is known, this presentation has various physical applications, e.g. the 
description of smectic liquid crystals \cite{Avi}, thin film blisters 
\cite{Gio},\cite{Ort} and convective pattern formation \cite{Erc}. 
Physically, such model can be regarded as the Landau model applied to a
system with vector order parameter. Some well-known Landau theories have
similar features. For example, the energy of a smectic-A liquid crystal 
has been described within such model \cite{Avi} where 
$\frac{\mathbf{\nabla}u(\mathbf{r})}{u(\mathbf{r})}$ represents the director 
field of the liquid crystal. The observed focal-conic defect structures can 
also be described in terms of this functional \cite{Gig}. The micro-magnetics give
one more example of the application of the functional (5). In particular, 
it can be used to describe magninization constrained by $|m| = 1$ within and 
$ m = 0$ outside the micro-magnetics \cite{Bal}. The unknown $ u(\mathbf{r})$ 
is purely curl-free, while $|m|$ just prefers to be divergence-free. Moreover, 
$|m|$ is restricted to unit vectors, while $ u(\mathbf{r})$ prefers to 
have unit magnitude. But the similarity should be clear, particularly for an 
isotropic ferromagnet \cite{Lan}. 

Another motivation to use the functional (4) arises from recent phenomenological
modeling of blisters in compressed thin ﬁlms \cite{Gio}. Early it has been 
suggested that the fold patterns of such blisters could be described by minimizing
the sum of membrane and bending energies \cite{Lan}. With some simpliﬁcation, this 
problem can be reduced to the free energy, given by (4). The same free energy can 
also be obtained for an equilibrium state of a free surface. In this case 
$ u(\mathbf{r})$ represents the height profile of the sheet (relative to a flat 
reference state). This model describes a fluctuating fluid membrane. 

The interpretation of the free energy appears also in the 
phase diffusion theory of pattern formation proposed by Cross and Newell 
\cite{Gros},\cite{New}. Of course, the physical justification of the 
models discussed above can be criticized \cite{Lob}. Nevertheless, they 
demonstrate the need to introduce new representations for the free energy 
functional in term of the order parameter. We have referred informally to 
the existence of an asymptotic variational problems. Now, let us consider
the possibility to generalize the representation of the free energy to the 
case when it depends not only on the order parameter,and its gradient, 
but also on their combination. Such generalization looks quite reasonable
if one bears in mind that, in the case of the functional with the first
order spatial derivative, the order parameter describes a phase transition
in a spatially homogeneous system, while the presence of the second-order 
derivative makes it possible to describe the formation of spatially 
inhomogeneous structures. Thus, we can expect that the combination of the
gradient terms with the scalar order parameter can be responsible for the
self-consistent influence of the order parameter on the spatial distribution 
and parameters of the ordered structures. In other words, the coupling between
the order parameter and its derivatives can influence possible deformations 
and confine possible inhomogeneous stable structures of the system.

So, let us postulate, that the order-parameter dependence of the free 
energy is given by
\begin{equation}
\begin{array}{cc}
f=al^{2}\left(\left|\mathbf{\nabla}\varphi(\mathbf{r})\right|\right)^{2}+b(\left(\left|\varphi(\mathbf{r})\right|\right)^{2}-1)^{2}+\\
+ml^{4}\left(\left|\mathbf{\nabla}\mathbf{\nabla}\varphi(\mathbf{r})\right|\right)^{2}+
n(l^{2}\left(\left|\mathbf{\nabla}\varphi(\mathbf{r})\right|\right)^{2}-1)^{2}+\\
+c\varphi^{2}(\mathbf{r})\left(l\mathbf{\nabla}\varphi(\mathbf{r})\right)^{2}
\end{array}
\end{equation}
where $l$ is the length of the order-parameter changing, $a$,$b$,$m$,$n$ 
and $c$ are parameters describing the influence of the gradient order parameter
and coupling between the order parameter and the appropriate derivative 
of this parameter. Hiving introduced the operator $D=l\mathbf{\nabla}$ we can
rewrite the free energy density as
\begin{equation}
\begin{array}{cc}
f=a\left(\left|D\varphi(\mathbf{r})\right|\right)^{2}+b(\left(\left|\varphi(\mathbf{r})\right|\right)^{2}-1)^{2}+\\
+m\left(\left|D^{2}\varphi(\mathbf{r})\right|\right)^{2}+
n(l^{2}\left(\left|D\varphi(\mathbf{r})\right|\right)^{2}-1)^{2}+\\
+c\varphi^{2}(\mathbf{r})\left(D\varphi(\mathbf{r})\right)^{2}
\end{array}
\end{equation}
If all the coefficients except $a$ and $b$ are equal to zero, we come to 
the standard theory of phase transitions. If $a$,$b$ and $c$ vanish wee come 
to the standard gradient theory. In the case of large values of $m$ ( $m$ is 
larger than other coefficients) we obtain the eikonal equation $\mathbf{\nabla}\varphi(\mathbf{r})=1$ 
which we supplement with the boundary conditions $\varphi(\mathbf{r})=0$ 
at the boundary.This eikonal equation has no smooth solution, but it has 
inﬁnitely many Lipschitz solutions. We can suggest that the energy can be
concentrated at the discontinuities of $\varphi(\mathbf{r})$. Thus, the 
singular part of the order parameter can provide a selection mechanism for    
the perturbations of the eiconal equation solutions which minimize the free energy.

In the general case, the minimum of the free energy satisfies the Euler-Lagrange 
equation:
\begin{equation}
 \frac{\delta f }{\delta \varphi(\mathbf{r})}=\frac{\partial f }{\partial \varphi(\mathbf{r})}-
D\frac{\partial f }{\partial D\varphi(\mathbf{r})}+ D^{2}\frac{\partial f }{\partial D^{2}\varphi(\mathbf{r})}
\end{equation}
which is reduced in our case to:
\begin{equation}
\begin{array}{cc}
mD^{4}\varphi(\mathbf{r})-(a-2n+6n(D\varphi(\mathbf{r}))^{2}+c\varphi^{2}(\mathbf{r}))D^{2}\varphi(\mathbf{r})\\-
c\varphi(\mathbf{r})(D\varphi(\mathbf{r}))^{2}-2b\varphi(\mathbf{r})(1-\varphi^{2}(\mathbf{r}))=0
\end{array}
\end{equation}
Let us consider in the one dimensional case some probable solutions of 
the Euler-Lagrange equation for various combinations of the coefficients introduced.

\textit{a) Linear solutions}

i) We shall look for a solution similar to the solution of the equation 
$D\varphi=-\varphi$ and $D^{2}\varphi=\varphi$. Substituting this  solution
in the Euler-Lagrange equation yields
\begin{equation}
 m\varphi-(a-2 n+c\varphi^{2}+6n\varphi^{2})\varphi-c\varphi^{3}-2b\varphi(1-\varphi^{2})=0
\end{equation}
This equation leads us to the following relations between the coefficients, 
$m-a=2(b-n)$ and $b-c=3m$. In the standard model of phase transitions $a=b=1$ 
and thus we have $m+2n=3$ and $c=1-3n$. If $m=n=1$ the coupling constant $c=-2$. 
If $a=1$ and $b=-1$ and $m=1$ and $n=-1$ the coupling constant $c=2$. 
In the case of the gradient presentation $a=b=0$ the exponential solution 
can be realized only for $n=1$, $m=-2$ and $c=3$. In the case of standard 
model with  $n=m=0$ the exponential solution exists for $b=1$, $a=-2$ and 
$c=1$. Thus, the exponential solution is realized for various combination 
of the introduced coefficients.

ii) Now let us look for another possible linear solution in the form of
a periodical function, namely $\varphi(\mathbf{r})=\varphi exp(ikr)$. 
This solution satisfies the equation $D^{2}\varphi(\mathbf{r})=-k^{2}\varphi(\mathbf{r})$.
Substituting of this solution into Euler-Lagrange equations leads to the relation 
\begin{equation}
\begin{array}{cc}
m k^{4}\varphi(\mathbf{r})+(a-2 n+c\varphi^{2}+6 n k^{2}\varphi^{2})k^{2}\varphi(\mathbf{r})\\-
c\varphi^{2}\varphi(\mathbf{r})-2 b\varphi(\mathbf{r})(1-\varphi^{2})=0
\end{array}
\end{equation}
which generates two conditions i.e. $mk^{4}+(a-2n)k^{2}-2b$ and
$6 n k^{2}+2b=0$, for which the spatial periodical distribution of the order 
parameter does not depend on the coupling constant. The conditions thus obtained 
can be used to find the wavelength of the periodic distribution of the order parameter.
The second condition yields $k^{2}=-\frac{b }{3n}$. On the other hand 
from the first condition one obtains $k^{2}=-\frac{a+4n}{m}$. This means that these
conditions are consistent with each other if $m b=3 n(a+4n)$. If $k^{2}=1$, when 
the period of the structure is equal to the size of the system under consideration,
we obtain  $m=-(a+4n)$ or $b=-3 n$. In the case of the standard theory of phase 
transitions with $a=b=1$,we obtain relation for the other coefficients i.e. 
$n=-\frac{1}{3}$ and $m=\frac{1}{3}$. It is important,that in the case under 
consideration the solution for the order parameter is independent of the coupling 
constant. The next step is to consider the realization of possible nonlinear 
solutions which satisfy the minimum of the free energy density.

\textit{a) Nonlinear solutions}

i)We start from the nonlinear solution of the standard model of phase transitions, 
namely, from the well-known the soliton solution which satisfies the equation 
\begin{equation}
 D^{2}\varphi=2\varphi(\varphi^{2}-1)
\end{equation}
The solution of this equation is $\varphi=tanh(lr)$. This solution satisfies 
an equation with the high-order derivatives that is given by $D\varphi=(1-\varphi^{2})$ 
and next relation $D^{4}\varphi=8\varphi(1-\varphi^{2})(3\varphi^{2}-1)$
Having substituted these relations into Euler-Lagrange equation one finds that
thus solution can be realized under the conditions $n=0$, $16m+2(a-b)-c=0$ and 
$c=8m$. In the case of the standard theory of phase transition $a=b$ all the other 
coefficients should be equal to zero. In the case of the generalized model we 
can observe the soliton solution for $a=1$,$b=2$, $c=2$ and $m=\frac{1}{4}$. 
We can also consider other nonlinear solutions of the standard model.

ii)The second example of the nonlinear solution of the standard model 
concerns the one $\varphi=sech(lr)$ that satisfied the nonlinear equation
\begin{equation}
 D^{2}\varphi=\varphi(1-2\varphi^{2})
\end{equation}
Assuming also that the equations $ D\varphi^{2}=\varphi^{2}(1-\varphi^{2})$ 
and $D^{4}\varphi=\varphi(1-20\varphi^{2}+24\varphi^{4})$ are satisfies
one obtains an equation for the coefficients of the generalized model.
In particular, the solution in the proposed form exists, if $m=a+2b$, $c=-8m$ 
and $a-c+b=10m$. This means, that in the case of the standard model $a=b$ does not
lead to the minimization of the generalized free energy functional.

iii)Finally, let us consider whether the solution $\varphi=ln ch (lr)$ can be
realized.This solution satisfies the equations $D^{2}\varphi=(1-(D\varphi^{2})$,
$D\varphi=tanh (lr)$ and $D^{4}\varphi=2(1-(D\varphi)^{2})(3(D\varphi)^{2}-1)$.
It follows from the Euler-Lagrange equation that this solution minimizes the free
energy functional under the conditions $a=b=c=0$, i.e., we obtain the well-
known form  
\begin{equation}
 f=(D^{2}\varphi)^{2}+(1-(D\varphi)^{2})
\end{equation}
The absolute minimum of the free energy is achieved for $\varphi=1$
and $D\varphi=1$, that cannot be satisfied. This means, that the
the eikonal solution satisfies the condition of the free energy divergence.
Such conditions can be realized only for spatially inhomogeneous order parameters. 

Thus, we have proposed a generalization of the standard and gradient theories
of the phase transitions by introducing the coupling between the order
parameter ant its gradient. Such generalization can be employed to describe
the phase transitions from spatially homogeneous to inhomogeneous states.
It is shown, that the solution of the standard and gradient models of phase
transitions can be inconsistent with the Euler-Lagrange equation generated
by the generalized functional of the free energy. In the general case
the requirement of consistency of the known solutions with the 
generalized description can be archived by the appropriate choice of 
the coupling between the order parameter and its gradient.

Specific examples considered in the present contribution shows, that
the structure formation observed experimentally can be described by various
phenomenological free energy functional which correspond to various 
sets of coefficients, i.e. there is no unique functional representation 
of the free energy related to the chosen spatially inhomogeneous
configuration of the order parameter. This uncertainty is generated
by the phenomenological description and could be eliminated in the
microscopic calculations.

This work was partially supported by the collaboration grant of the
National Academy of Science of Ukraine and the Russian 
Foundation of Fundamental Researches.

\end{document}